\documentclass[twocolumn,secnumarabic,nobibnotes, aps, pre]{revtex4-1}
\usepackage{graphicx,amsmath,amssymb,commath,mathtools,braket}
\usepackage{hyperref}
\usepackage{cleveref}

\usepackage{todonotes}

\crefname{equation}{Eq.}{Eqs.}
\crefname{figure}{Fig.}{Figs.}
\crefname{appendix}{Appendix.}{Appendices.}
\crefname{section}{Sec.}{Secs.}

\newcommand{\Average}[1]{\DegreeNodeSum \Measure \WeightSum{#1}}
\newcommand{\AverageReduced}[1]{\DegreeNodeSum \MeasureReduced \WeightSum{#1}}
\newcommand{\DegreeNodeSum}{\sum_k \frac{p_k k}{\MeanK}}
\newcommand{\DegreeSum}{\sum\limits_k p_k}
\DeclareMathOperator{\For}{for}
\newcommand{\Hamiltonian}{h^{(k)}}
\newcommand{\HamiltonianComp}[1]{h^{(k)}_{#1}}
\newcommand{\HamiltonianFull}{\mathcal{H}^{(k)}}
\newcommand{\MeanK}{\Braket{k}}
\newcommand{\Measure}{\iint \mathcal{D} z \mathcal{D}w\,}
\newcommand{\MeasureReduced}{\int \mathcal{D} z\,}
\newcommand{\MeasureRegion}[1]{\iint\limits_{#1} \mathcal{D}z \,}
\newcommand{\Modularity}{Q_\textrm{MOD}}
\newcommand{\ModularityMax}{\Modularity^{*}}
\newcommand{\ModularityMaxAvr}{[\ModularityMax]_c}
\newcommand{\ModularityMaxEta}{[ Q_{\textrm{MOD},\eta}^{*} ]_c}
\newcommand{\Order}[1]{\mathcal{O} \left( #1 \right) }
\newcommand{\OrderLBefore}{\overline{L}_{\mu\nu}^{\alpha}}
\newcommand{\OrderMBefore}{\overline{M}_{\mu}^{\alpha}}
\newcommand{\OrderQBefore}{\overline{Q}_{\mu\nu}^{\alpha\beta}}
\newcommand{\PottsSpin}{\delta(\sigma_i^\alpha, \sigma_j^\alpha)}
\newcommand{\Spin}[2]{S_{#1}^{#2}}
\newcommand{\SpinAsymm}{\mathcal{A}_{\mu\nu}}
\newcommand{\SpinSymm}{\mathcal{S}_{\mu\nu}}
\newcommand{\SpinVec}[2]{\vec{S}_{#2}^{(#1)}}
\newcommand{\SpinVecComp}[2]{S_{#1, #2}}
\newcommand{\Tr}[1]{\mathrm{Tr}_{#1}}
\newcommand{\WeightSum}[1]{\Braket{#1}_{\Hamiltonian}}

\begin{document}
\setcounter{page}{1}
\title{Ground state energy of $q$-state Potts model: the minimum modularity}
\author{J. S. Lee$^{1}$}\email{jslee@kias.re.kr}
\author{S. Hwang$^{2}$, J. Yeo$^{3}$, D. Kim$^{1}$}
\author{B. Kahng$^{2}$}\email{bkahng@snu.ac.kr}
\affiliation{{$^1$School of Physics, Korea Institute for
Advanced Study, Seoul 130-722, Republic of Korea}\\
{$^2$Department of Physics and Astronomy, Seoul National
University, Seoul 151-747,Korea}\\
{$^3$}School of Physics, Konkuk University, Seoul 143-701, Korea}
\date{\today}

\begin{abstract}
A wide range of interacting systems can be described by complex networks. A common feature of such networks is that they consist of several communities or modules, the degree of which may quantified as the \emph{modularity}.
However, even a random uncorrelated network, which has no obvious modular structure, has a finite modularity due to the quenched disorder.
For this reason, the modularity of a given network is meaningful only when it is compared with that of a randomized network with the same degree distribution. 
In this context, it is important to calculate the modularity of a random uncorrelated network with an arbitrary degree distribution. 
The modularity of a random network has been calculated [Phys. Rev. E \textbf{76}, 015102 (2007)]; however, this was limited to the case whereby the network was assumed to have only two communities, and it is evident that the modularity should be calculated in general with $q(\geq 2)$ communities.
Here, we calculate the modularity for $q$ communities by evaluating the ground state energy of the $q$-state Potts Hamiltonian, based on replica symmetric solutions assuming that the mean degree is large. 
We found that the modularity is proportional to $\langle \sqrt{k} \rangle / \MeanK$ regardless of $q$ and that only the coefficient depends on $q$. 
In particular, when the degree distribution follows a power law, the modularity is proportional to $\MeanK^{-1/2}$. 
Our analytical results are confirmed by comparison with numerical simulations. Therefore, our results can be used as reference values for real-world networks.
\end{abstract}
\pacs{05}

 \maketitle


\section{Introduction}

A wide range of networks, including, for example, the Internet, the world wide web, social relationships, and biological systems~\cite{Eriksen2003,Eckmann2002,Arenas2004,Holme2003}, may appear unrelated to each other. However, it has recently been shown that there exist several common features in such networks, including the existence of hub and fat-tailed degree distributions~\cite{Amaral2000,Albert2002,Barabasi1999}.
In particular, one important common feature is that a network consists of several \emph{communities}, which are densely connected sub-networks compared with other parts of the network.

Understanding the community structure of a given network is of practical importance. A set of nodes in the same community typically has similar properties or functions. For example, nodes belonging to the same community found in the world wide web~\cite{Flake2002} and social networks~\cite{Girvan2002} have similar topics and identities, respectively.
In addition, nodes in the same community of a metabolic network have been shown to have similar metabolic functions~\cite{Holme2003,Guimera2005}. Therefore, identifying the community structure provides information that aids in the understanding of the role of a specific node in a network.
Moreover, the analysis of community structures of gene-disease and
metabolite-disease networks may provide a method to predict complications associated with diseases~\cite{Goh2007}. 

Motivated by such practical importance, many authors have attempted to identify the optimal community structure of a given network, and a number of sophisticated algorithms to detect the possible optimal community structure have been reported~\cite{Newman2004, Wu2004,Radicchi2004, Newman2004a,Fortunato2004,Reichardt2004, Donetti2004,Zhou2004,Newman2006a,Danon2005,Danon2006,Reichardt2006}.
Most of these algorithms make use of the \emph{property} that the link density within a community is much larger than the inter-community link density. 
Therefore, it is crucial for community-detection algorithms to employ a suitable function to quantify such a \emph{property}. A widely used function for this purpose is the \emph{modularity}, introduced by Newman and Girvan~\cite{Newman2004}. The modularity function takes a community configuration as its argument and returns a value between $0$ and $1$.
The modularity represents how modular a given network is, {\em i.e.}, a larger modularity corresponds to a network that is more modularized or has a richer community structure.

\begin{figure}
\resizebox{0.49\columnwidth}{!}{\includegraphics{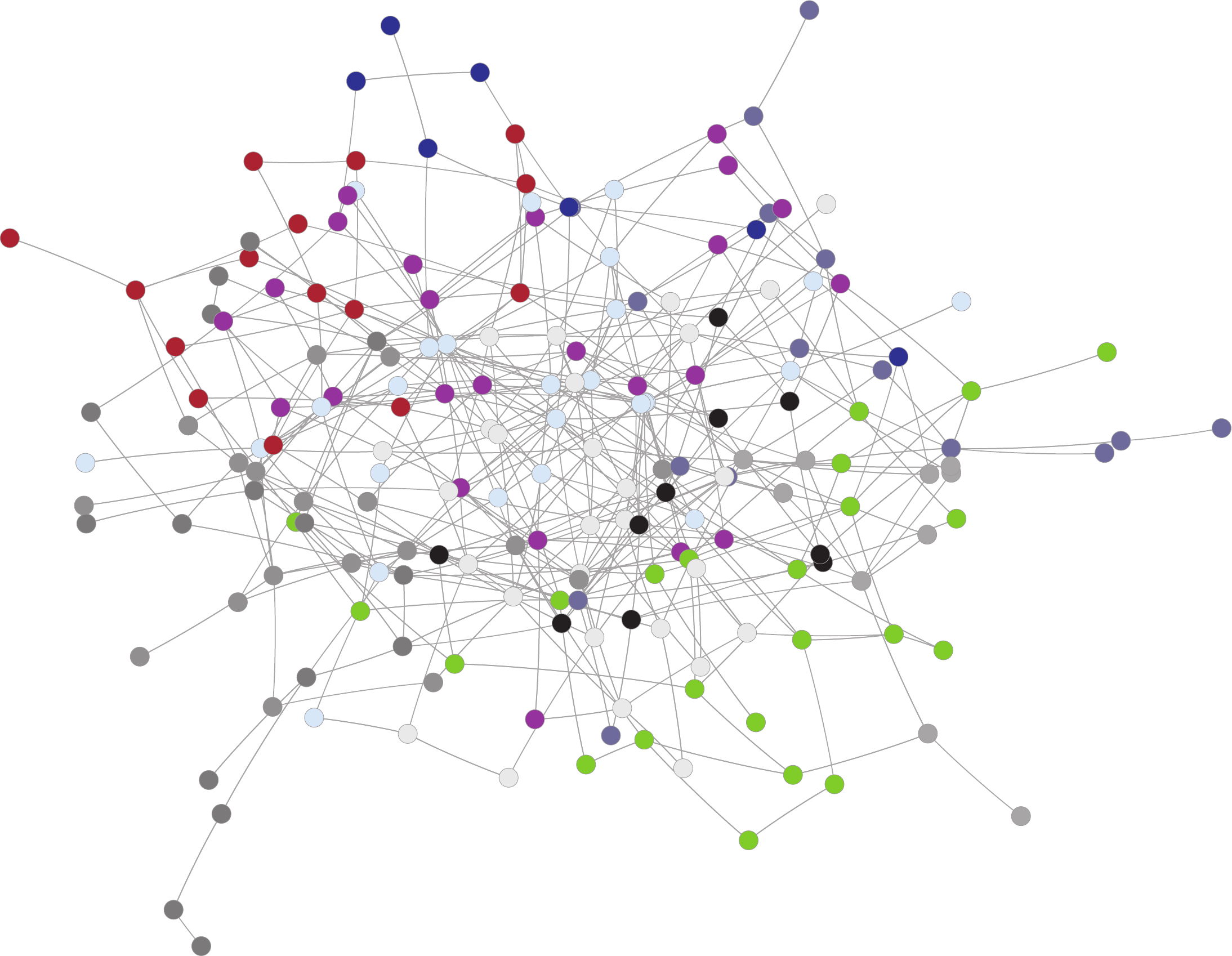}}
\resizebox{0.49\columnwidth}{!}{\includegraphics{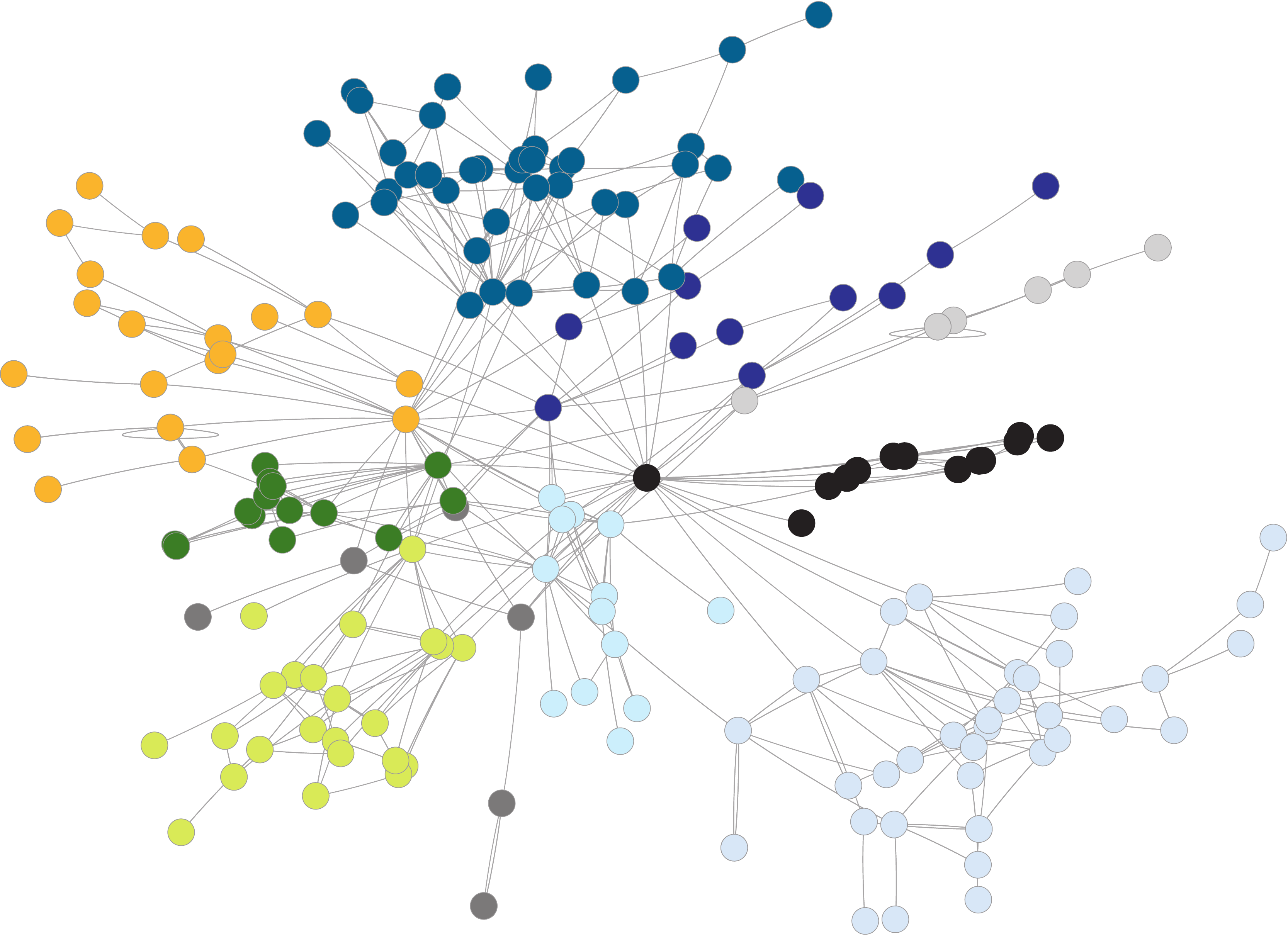}}
\caption{[color online] Examples of a random uncorrelated and a modular network.
Each color represents a different community, as identified by
	the $q$-state Potts model.
(a) A static model with a modularity of 0.51. (b) A modular network with a modularity of 0.72.} \label{fig1}
\end{figure}

The absolute value of the modularity, however, is not necessarily helpful in discerning how modular a network is.
In other words, a finite modularity does not guarantee a truly modular structure of a network.
In Ref.~\cite{Guimera2004}, Guimer\`a \emph{et al.}~showed that even a random uncorrelated network, which presumably does not have a modular structure, has a finite modularity because of the presence of quenched disorder.
For example, \cref{fig1}(a) shows a random uncorrelated network generated using a static model~\cite{Goh2001}.
Despite the lack of any obvious community structure, the modularity of this network is $0.51$, which may be considered to be a relatively large value of the modularity in the usual sense.
\cref{fig1}(b) shows another network with the same size and the same degree distribution. In this case, we can see a clear community structure, and the modularity is $0.72$, which is larger than that of the first example.

It follows that the modularity is meaningful only when compared with a random uncorrelated network with the same degree distribution.
Therefore, calculating the modularity of random uncorrelated networks with an arbitrary degree distribution is important to determine a reference modularity.
Reichardt \emph{et al.}~\cite{Reichardt2007} found that calculating the ground state energy of an Ising model of a network is equivalent to finding the modularity of the network if the network has two communities.
Using this equivalence, they calculated the modularity of a random uncorrelated network with an arbitrary degree distribution assuming that the network had only two communities.

In general, however, it is clear that the modularity should be calculated with an arbitrary number of communities. Here, we denote the number of communities as $q(\geq 2)$, and we calculate the modularity of networks with $q$ communities. 
To achieve this, we map the modularity function for a network with $q$ communities onto the ground state energy of the $q$-state Potts model. 
We then calculate the energy of the Potts model for a random uncorrelated network with an arbitrary degree distribution
in the large mean-degree limit. Our main result is that the ground state energy is given by $C(q) \Braket{\sqrt{k}}/\MeanK$, \cref{modularity_solution}, where the coefficient $C(q)=- \sqrt{\frac{2 - \frac{1}{q-1}}{2 \pi}}$ and $\MeanK$ is the mean degree of the network. Note that only the coefficient $C(q)$ is $q$-dependent, and approaches a finite value when $q\rightarrow \infty$.  For a scale-free network, $\langle \sqrt{k} \rangle/\MeanK \propto \MeanK^{-1/2}$. 

The remainder of this paper is organized as follows.
In \cref{sec:GroundStateVSModularity}, we first describe how the problem of finding a community structure can be mapped to that of finding the ground state of the $q$-state Potts model. This is achieved by comparing the modularity function with the Hamiltonian of the $q$-state Potts model. We then derive the replica-symmetric solutions for the free energy and energy of the Hamiltonian.
In \cref{sec:GroundStateEnergy}, we give analytic expressions for the energy, especially the ground-state energy, for several $q$. We also provide a conjecture for the ground state energy of the Hamiltonian for an arbitrary $q$.
In \cref{sec:Numerical}, we compare the analytical results with numerical simulations.

\section{Analytic solutions for the $q$-state Potts model}
\label{sec:GroundStateVSModularity}
\subsection{Hamiltonian of the $q$-state Potts model}
\label{sec:Potts_model}
We begin by describing the modularity and discussing how it is related to the $q$-state Potts model.
Consider a network composed of $N$ nodes, $L$ edges, and $q$ communities. The degree distribution of the network is $p_k$.
Let us arbitrarily assign a unique integer in the range from $1$ to $q$ to each community.
Then let $\sigma_i$ denote the number of communities assigned to a node $i$.
The modularity $\Modularity$~\cite{Newman2006} is defined as the difference between the proportion of the intra-community edges of a given network and
the expected proportion of such edges in a random uncorrelated network with the same degree distribution.
That is, $\Modularity$ is given by
\begin{eqnarray}
\Modularity
	&=&\frac{1}{L}(\textrm{number of intra-community edges})\nonumber \\
	& &-\frac{1}{L}(\textrm{expected number of such edges})\nonumber \\
	&=&\frac{1}{L}
		\sum_{i<j} \left(
			A_{ij} - \frac{k_i k_j}{\langle k \rangle N}
		 \right) 
		\delta (\sigma_i ,\sigma_j),
\end{eqnarray}
where the adjacency matrix element $A_{ij}=1$ if there is an edge between two distinct nodes $i$ and $j$; otherwise, $A_{ij} = 0$. Here, $k_i$ denotes the degree of node $i$, {\emph i.e.}, $k_i=\sum_j A_{ij}$, and $\MeanK$ is the mean degree of the network.
Note that the term $k_i k_j/(\MeanK N)\equiv f_{ij}$ in the above expression is the connection probability between nodes $i$ and $j$ in a random uncorrelated network.

If a specific community structure $\{ \sigma_1, \cdots, \sigma_N \}$ is initially given, the calculation of the modularity is straightforward.
However, in most cases, this information is not known \emph{a priori}; rather, the optimal community structure is determined as the one that maximizes the modularity, which is chosen from all possible configurations of $\{ \sigma_1,\cdots, \sigma_N \}$. This maximum modularity will be denoted by $\Modularity^*$.
Therefore, a major task for community detection is finding the community configuration that maximizes the modularity.
However, since the number of all possible configurations increases exponentially with $N$ ($\sim q^N$), 
it is not generally feasible to enumerate and test all of them for a network with large $N$.
	
To avoid such difficulties, several \emph{feasible} algorithms~\cite{Guimera2004,Guimera2005a,Massen2005} have been proposed.
One particularly interesting approach is to use the $q$-state
Potts model, the Hamiltonian of which is given by~\cite{Reichardt2007}
\begin{equation}
		\mathcal H = -\frac{1}{\MeanK}\sum_{i<j} (A_{ij}-\eta f_{ij})
	\delta(\sigma_i,\sigma_j).~~~\left( f_{ij}=\frac{k_i
	k_j}{\MeanK N} \right)
	\label{PottsH}
\end{equation}
where $\sigma_i$ denotes the spin state of node $i$ of $q$ possible spin states and $\eta$ is a control parameter.
Note that the connection probability $f_{ij}$ is typically very small, {\emph i.e.}, $f_{ij}\ll 1$.
Therefore, when $A_{ij}=1$ ($A_{ij}=0$), the coupling constant between nodes $i$ and $j$ becomes positive (negative);
thus, two spins, $\sigma_i$ and $\sigma_j$, tend to be in the same (different) spin state(s) in order to lower the energy $E$ of the Hamiltonian.
The ground-state energy $E_g$ of this model can be obtained from the spin configuration by minimizing the Hamiltonian.
When $\eta=1$, the ground-state energy is proportional to the maximized modularity $\Modularity^*$ {\emph i.e.},
\begin{equation}
	\Modularity^* = -2 E_g/N.
	\label{QErelation}
\end{equation}
Therefore, finding the community structure of a network now becomes a problem of searching the ground state of the $q$-state Potts model Hamiltonian.

\subsection{Free energy}
In this section, we describe the calculation of the free energy of the $q$-state Potts model Hamiltonian~(\ref{PottsH}) for an uncorrelated random network with an arbitrary degree distribution $p_k$  as a reference value for the modularity. 
We assume that the typical free energy of \cref{PottsH} is the same as the quenched average of the free energy over the network configurations $\{A_{ij} \}$.
Using the replica method, the configuration-averaged free energy is given by
\begin{equation}
	[\ln Z]_c = \lim_{n\rightarrow 0} \frac{ [ Z^n ]_c - 1}{n}.
	\label{replica}
\end{equation}
where $Z$ is the partition function of the Hamiltonian for a one-network configuration and $[ \cdots]_c$ denotes the configuration-ensemble average \cite{Sherrington1975,Nishimori2011}. 
In the context of the replica method, $n$ is assumed to be a non-zero integer, prior to discussing the limit $n=0$.
For any integer $n$, we can write the above expression as
\begin{align*}
&[Z^n]_c = \left[( \Tr{i}~ e^{-\beta {\mathcal H}})^n \right]_c \\
	&= \int \prod_{i<j} \dif J_{ij} P(J_{ij}) \Tr{i,\alpha}
	\exp \left [
		\frac{\beta}{\MeanK} \sum_{\substack{i<j \\\alpha}}  J_{ij}
		\PottsSpin
	\right ],
\end{align*}
where $\beta$ is the inverse temperature, $J_{ij}\equiv(A_{ij}-\eta f_{ij})$, and 
$\Tr{i, \alpha}$ denotes the sum of all possible spin states $\sigma_i^\alpha$ over all nodes in all the replicas.
Using $P(J_{ij}) = f_{ij} \delta(J_{ij} -1+ \eta f_{ij}) + (1-f_{ij}) \delta(J_{ij}+\eta f_{ij})$, $[Z^n]_c$ becomes
\begin{widetext}
\begin{equation}
	[Z^n]_c = \Tr{i,\alpha} \exp\left(
		-\frac{\beta}{\MeanK} \sum_{i<j} \sum_\alpha \eta f_{ij} \PottsSpin
	\right)
	\exp\left[
		\sum_{i<j} \ln\left(
			 1 + f_{ij} \Set{
			 	\exp\left[
			 		\frac{\beta}{\MeanK} \sum_{\alpha} \PottsSpin
				\right] - 1 
			} 
		\right) 
	\right]
\end{equation}
Now, we use an approximation
\begin{equation}
	\sum_{i<j} \ln ( 1+f_{ij} D_{ij} ) \approx \sum_{i<j} f_{ij} D_{ij} = \sum_{i<j} \frac{k_i k_j}{\MeanK N} D_{ij},
\end{equation}
which is valid in the thermodynamic limit for a wide range of uncorrelated ensembles~\cite{Kim2005}. 
Then, $[Z^n]_c$ becomes
\begin{align}
	[Z^n]_c&= \Tr{i,\alpha} \exp\left[
		\sum_{i<j}\frac{k_i k_j}{\MeanK N}
		\left(
			-\frac{\beta \eta}{\MeanK} \sum_\alpha \PottsSpin
			+ \exp\left(
				\frac{\beta}{\MeanK} \sum_\alpha \PottsSpin
			\right)-1 
		\right)
	\right].
	\label{derive1}
\end{align}
\end{widetext}

\begin{figure}
\resizebox{0.99\columnwidth}{!}{\includegraphics{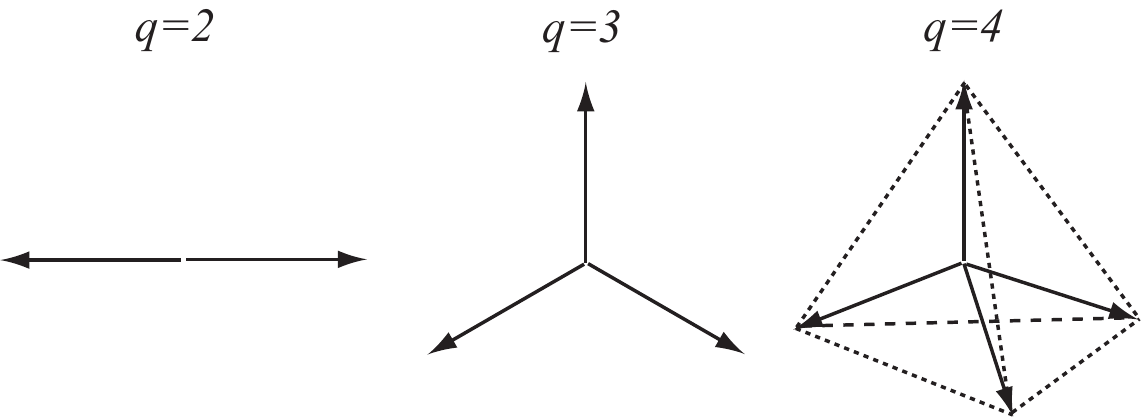}}
\caption{Potts spin vector. $q$-state Potts spin can be mapped into vertices of a $r(=q-1)$-dimensional simplex.} \label{fig:PottsSpin}
\end{figure}

To manipulate the Kronecker delta function, it is convenient to adopt the vector representation for $q$-state Potts spins~\cite{Wu1982,Lee2004}.
As shown in \cref{fig:PottsSpin}, each $q$-state Potts spin $\sigma_i$ can be mapped to a $q$-1 dimensional vector $\vec{S_i}$. The angle between any two vectors is identical.
Then, the Kronecker delta function can be written as
\begin{align}
	\delta(\sigma_i,\sigma_j) &= \frac{1}{q} 
		(r \vec{S}_i \cdot \vec{S}_j +1) \nonumber\\
	&= \frac{1}{q}(r\sum_\mu S_{i\mu}S_{j\mu} +1),
\label{delta}
\end{align}
where 
\begin{equation}
	r \equiv q-1.
	\label{r}
\end{equation}
The vector-component index $\mu$ varies from $1$ to $r$. 

In this work, we consider the densely connected limit~\cite{Fu1986}, i.e., $\beta \ll \MeanK$ for fixed $\beta$. Then, by expanding the exponential term $\exp \left(\frac{\beta}{\MeanK} \sum_\alpha \PottsSpin \right)$ in Eq.~(\ref{derive1}) up to the second order in $\frac{\beta}{\MeanK}$
	and by using Eq.~(\ref{delta}), \cref{derive1} can be written as
\begin{equation}
[Z^n]_c = \exp\left( \frac{n\beta(1-\eta) N }{ 2q}
	\right) \Lambda
	\label{PartitionFunction}
\end{equation}
with
\begin{widetext}
\begin{equation}
	\Lambda = \textrm{Tr}_{i,\alpha} \exp\left[
	\frac{C_1 N}{2 } \sum_{\alpha, \mu}
	\left(\sum_i \frac{k_i}{\MeanK N} S_{i\mu}^\alpha \right)^2  + \frac{C_2 N}{2}
	\sum_{\alpha\neq\beta,\mu\nu} \left( \sum_i \frac{k_i}{\MeanK N} S_{i\mu}^{\alpha}
	S_{i\nu}^\beta \right)^2 + \frac{C_2 N}{2}
	\sum_{\alpha,\mu\nu} \left( \sum_i \frac{k_i}{\MeanK N} S_{i\mu}^{\alpha}
	S_{i\nu}^\alpha \right)^2\right],
\end{equation}
where 
\begin{equation}
	C_1 = \frac{\beta (1-\eta) r}{q },
	~~~ C_2 = \frac{\beta^2 r^2}{2\MeanK q^2}.
	\label{coeff}
\end{equation}
Note that the terms which are higher order than $n$ are ignored in the above derivation since they will vanish as $n\to 0$. 

Now, performing the Hubbard-Stratonovich transform on each quadratic term in $\Lambda$ and applying the saddle point method subsequently, $\Lambda$ becomes
\begin{eqnarray}
	\Lambda &=& \exp \left[
		- \frac{C_1 N}{2} \sum_{\alpha\mu} (\OrderMBefore)^2
		- \sum_{\alpha\neq\beta,\mu\nu} \frac{C_2N}{2} (\OrderQBefore)^2
		- \sum_{\alpha,\mu\nu}\frac{C_2 N}{2} (\OrderLBefore)^2
		+ N \DegreeSum \ln \Tr{\alpha} \exp \HamiltonianFull
	\right], 
\end{eqnarray} 
where $p_k$ is the degree distribution of a given network and $\HamiltonianFull$ is given by,
\begin{equation}
\HamiltonianFull = \frac{k}{\MeanK}\left(
	C_1 \sum_{\alpha\mu} S_{\mu}^\alpha \OrderMBefore
	+ C_2 \sum_{\alpha\neq\beta,\mu\nu} S_{\mu}^{\alpha} S_{\nu}^\beta \OrderQBefore
	+ C_2 \sum_{\alpha,\mu\nu} S_{\mu}^{\alpha} S_{\nu}^\alpha \OrderLBefore
\right).
\end{equation}
\end{widetext}
Here, $\OrderMBefore$, $\OrderQBefore$ and $\OrderLBefore$ are chosen to satisfy the saddle point condition. Their explicit replica-symmetric forms will be shown later in \cref{SelfEqM,SelfEqL,SelfEqQ}.

At this stage, we seek a replica symmetric solution so that we assume
	$\OrderMBefore \to M_\mu$,
	$\OrderQBefore \to Q_{\mu\nu}$ and
	$\OrderLBefore \to L_{\mu\nu}$.
Then, $\Lambda$ and $\HamiltonianFull$ can be simplified as,
\begin{align}
\Lambda = \exp 
&\left[
	-\frac{N C_1 n}{2} \sum_{\mu}  (M_\mu)^2
	-\frac{N C_2 n(n-1)}{2}  \sum_{\mu\nu}  (Q_{\mu\nu})^2  
\right. \nonumber \\
&\left.
	-\frac{N C_2 n}{2} \sum_{\mu\nu}  (L_{\mu\nu})^2  
	+ N \DegreeSum \ln \Tr{\alpha} \exp \HamiltonianFull
\right] \nonumber \\
\label{replicaSymSol}
\end{align}
and
\begin{align}
\frac{\MeanK}{k} \HamiltonianFull &= C_1 \sum_{\alpha \mu} \Spin{\mu}{\alpha} M_\mu \nonumber \\
	& + C_2 \sum_{\alpha} \sum_{\mu \nu} \Spin{\mu}{\alpha} \Spin{\nu}{\alpha}
		\left(L_{\mu\nu} - Q_{\mu \nu} \right) \nonumber \\
	& + C_2  \sum_{\mu \nu} Q_{\mu\nu} 
		\left( \sum_{\alpha } \Spin{\mu}{\alpha} \right)
		\left( \sum_{\beta } \Spin{\nu}{\beta} \right) .
\label{replicaSymHamiltonian}
\end{align}
The quadratic nature of the last term in \cref{replicaSymHamiltonian} allows us 
	to perform the modified Hubbard-Stratonovich transform. 
In \cref{appendix:Hubbard}, it is shown that
\begin{align}
\ln \Tr{\alpha} \exp \HamiltonianFull 
= n \Measure \ln \Tr{} \exp \Hamiltonian + \Order{n^2},
\label{LogHamiltonian}
\end{align}
where $\mathcal{D} z = \prod_{\mu\nu} \frac{\dif z_{\mu \nu}}{\sqrt{2 \pi}} \exp\left(-\frac{z_{\mu \nu}^2}{2} \right) $ and
$\mathcal{D}  w  = \prod_{\mu\nu}  \frac{\dif w_{\mu \nu}}{\sqrt{2 \pi}} \exp\left( -\frac{w_{\mu \nu}^2}{2} \right) $ and $\Hamiltonian$ is defined as
\begin{align}
\Hamiltonian \equiv& 
	\frac{k C_1}{\MeanK} \sum_{\mu}  \Spin{\mu}{} M_{\mu} \nonumber \\
	+& \frac{k C_2}{\MeanK}  \sum_{\mu\nu} \Spin{\mu}{} \Spin{\nu}{} \left(L_{\mu\nu}- Q_{\mu \nu} \right) \nonumber \\
	+& \sum_{\mu\nu}  \sqrt{\frac{2 k C_2 Q_{\mu\nu}}{\MeanK}} 
	\left\{
		\SpinSymm z_{\mu\nu}
		+ \SpinAsymm w_{\mu\nu}
	\right\},
\label{HamiltonianK}
\end{align}
with $\SpinSymm \equiv \frac{1}{2} (\Spin{\mu}{} + \Spin{\nu}{})$ and $\SpinAsymm \equiv \frac{i}{2} (\Spin{\mu}{} - \Spin{\nu}{})$.

From \cref{PartitionFunction,replicaSymSol,LogHamiltonian}, the free energy density is given by,
\begin{align}
f &= - \frac{1}{\beta} \lim_{n \to 0} \dfrac{ [Z^n] - 1}{nN} \nonumber\\
&= - \frac{(1-\eta)}{2q}
	+\frac{ C_1 }{2\beta} \sum_\mu M_\mu^2 
	+\frac{C_2}{2\beta} \sum_{\mu\nu} \left( L_{\mu\nu}^2 - Q_{\mu\nu}^2 \right) \nonumber \\ 
& - \frac{1}{\beta} \DegreeSum \Measure \ln 
\left(
	\Tr{} \exp \Hamiltonian
\right).
\label{FreeEnergy}
\end{align}
Here, $M_\mu$, $L_{\mu\nu}$, and $Q_{\mu\nu}$ are determined by minimization of the free energy. For $M_\mu$, the condition $\dpd{f}{M_\mu} =0$ gives
\begin{align}
M_\mu = \Average{S_\mu}
\label{SelfEqM}
\end{align}
where $\WeightSum{(\bullet)}$ denotes the expectation value with respect to $\Hamiltonian$,
	namely, $\WeightSum{(\bullet)} \equiv \frac{\Tr{} (\bullet) \exp  \Hamiltonian}{\Tr{} \exp \Hamiltonian}$.
Similarly, one can find
\begin{align}
L_{\mu \nu} = \Average{S_{\mu} S_{\nu}}
\label{SelfEqL}
\end{align}
and 
\begin{align}
Q_{\mu \nu} =& L_{\mu \nu}
- \DegreeSum \sqrt{\frac{k }{2\MeanK C_2 Q_{\mu \nu}}}  \nonumber \\
&	\Measure \WeightSum{  
	\SpinSymm z_{\mu\nu}
	+ \SpinAsymm w_{\mu\nu}
	} \nonumber \\
=&
\DegreeNodeSum \Measure 
	\WeightSum{ S_{\mu} } 
	\WeightSum{ S_{\nu} },
\label{SelfEqQ}
\end{align}
where the last equality in \cref{SelfEqQ} is obtained by integration by parts.
From \cref{SelfEqL,SelfEqQ}, one can easily check that $L_{\mu\nu} = L_{\nu\mu}$ and $Q_{\mu\nu} = Q_{\nu\mu}$.
By a proper rotation of $r$-dimensional space, any $r$-dimensional vector ($M_1$, $M_2$,$\cdots$, $M_r$) can be transformed into one satisfying the following condition,
\begin{align}
M_\mu = M_1 \delta_{\mu 1}.
\label{MDirection}
\end{align}
In this coordinate setting, one can prove some important identities for $Q_{\mu\nu}$ and $L_{\mu \nu}$ such as $Q_{\mu \nu} = L_{\mu \nu} =0$ for $\mu \ne \nu$, $L_{\mu \mu} = L_{\nu \nu}$ and $Q_{\mu \mu} = Q_{\nu \nu}$ for $\mu >1$ and $\nu >1$, and $\sum_{\mu=1}^{q} L_{\mu \mu} =1$, which are derived in \cref{appendixC}.
Using them, we finally obtain
\begin{align}
f &= - \frac{(1-\eta)}{2q}
	+\frac{ C_1 }{2\beta}  M_1^2 
	+\frac{C_2}{2\beta} \sum_{\mu} \left( L_{\mu\mu}^2 - Q_{\mu\mu}^2 \right) \nonumber \\ 
& - \frac{1}{\beta} \DegreeSum \MeasureReduced \ln 
\left(
	\Tr{} \exp \Hamiltonian
\right),
\label{FreeEnergy1}
\end{align}
where $\Hamiltonian$ is now simplified as \cref{HamiltonianKReduced}.
Note that in the above equation the integral with respect to $\int\mathcal{D} w$ disappears and $\mathcal{D} z$ is reduced to $\prod_{\mu} \frac{\dif z_{\mu \mu}}{\sqrt{2 \pi}} \exp\left(-\frac{z_{\mu \mu}^2}{2} \right)$ (see \cref{appendixC}).
From now on, $\MeasureReduced$ means the product of integrals with respect to only the diagonal integral variables $z_{\mu\mu}$ if there is no other comment.

\subsection{Energy}

From \cref{FreeEnergy1}, the energy $E$ is given by 
\begin{widetext}
\begin{align}
E/N = \dpd{(\beta f)} {\beta } &= - \frac{(1-\eta)}{2q}
	+\frac{ C_1 }{2\beta}  M_{1}^2
	+\frac{C_2}{\beta} \sum_{\mu} \left( L_{\mu\mu}^2 - Q_{\mu\mu}^2 \right) \nonumber \\ 
& - \DegreeSum \MeasureReduced \WeightSum{
	\frac{k C_1}{\beta \MeanK}  \Spin{1}{} M_{1}
	+ \frac{2 k C_2}{\beta \MeanK}  \sum_{\mu} \Spin{\mu}{2} \left(L_{\mu\mu} - Q_{\mu\mu} \right)
	+  \sum_{\mu}  \sqrt{\frac{2 k C_2 Q_{\mu\mu}}{\beta^2 \MeanK}} \Spin{\mu}{} z_{\mu\mu}
	}.
\label{EnergyIntermediate}
\end{align}
\end{widetext}
Using \cref{SelfEqM,SelfEqL,SelfEqQ}, the three terms in the average $\WeightSum{(\bullet)}$ in \cref{EnergyIntermediate} can be reduced to
\begin{subequations}
\begin{align}
\AverageReduced{\Spin{1}{}} &= M_{1} \label{SelfConsistent1} \\
\AverageReduced{\Spin{\mu}{2}} &= L_{\mu\mu} \label{SelfConsistent2} \\
\DegreeSum \MeasureReduced \WeightSum{  \frac{\sqrt{k}S_\mu z_{\mu\mu} }{\sqrt{\beta^2 \MeanK}}  } 
&= \frac{ \sqrt{2 C_2 Q_{\mu\mu}}}{\beta} \left( L_{\mu\mu} - Q_{\mu\mu} \right). \label{SelfConsistent3}
\end{align}
\label{SelfConsistent} 
\end{subequations}
With these equations, one can simplify \cref{EnergyIntermediate} as
\begin{align}
E/N \ 
=& - \frac{(1-\eta)}{2q}
	-\frac{ C_1 }{2\beta} M_{1}^2
	-\frac{C_2}{\beta} \sum_{\mu} \left( L_{\mu\mu}^2 - Q_{\mu\mu}^2 \right).
\label{Energy}
\end{align}

By plugging the solutions of the self-consistent equations~(\ref{SelfConsistent}) into \cref{Energy},
	we can calculate the energy for any $\beta$ and $\eta$.

\subsection{Energy for $\eta=1$ case}
\label{detail}

When $\eta = 1$, the energy is proportional to the modularity~\cite{Reichardt2007,Newman2004} as explained in \cref{sec:Potts_model}.
In this case, the order parameter $M_{1}$ becomes zero, $L_{\mu\mu} = \frac{1}{r}$ for all $\mu$ as shown in \cref{OrderLFinFirst,OrderLFin}, and  $Q_{\mu\mu}=Q_{\nu\nu}$ for any $\mu$ and $\nu$ as shown in \cref{eta1caseConditionQ}.
Thus, \cref{Energy} is simplified as
\begin{align}
E/N =& -\frac{C_2}{\beta} \sum_{\mu} \left( L_{\mu\mu}^2 - Q_{\mu\mu}^2 \right) = -\frac{r C_2}{\beta}  \left( L_{11}^2 - Q_{11}^2 \right).
\label{EnergyEtaOneIntermediate}
\end{align}
As $\beta$ becomes large, \cref{SelfEqQ} implies that $Q_{\mu\mu} = L_{\mu\mu} + \Order{1/\beta}$. The sub-leading order term becomes
\begin{align}
\beta(Q_{\mu \mu} - L_{\mu \mu}) = 
- \DegreeSum \sqrt{\frac{k q^2}{r}} \MeasureReduced z_{\mu\mu} \WeightSum{
	\Spin{\mu}{} 
}.
\end{align}
Finally, one can obtain
\begin{align}
E/N = - \frac{r \sqrt{r}}{\MeanK q} \DegreeSum \sqrt{k} \MeasureReduced z_{\mu\mu} \WeightSum{
	\Spin{\mu}{} 
}.
\label{EnergyEtaOneBetaInfinite}
\end{align}

\section{Ground state Energy for each $q$}
\label{sec:GroundStateEnergy}

\subsection{$q=2$}
In this case, Potts spins become one-dimensional vectors, which greatly simplifies the trace with respect to $\Hamiltonian$ as follows,
\begin{align}
\Tr{} \exp \Hamiltonian 
= \exp \left(
	\frac{\beta^2 k}{8 \MeanK^2} ( 1-Q_{11} ) 
\right)
2 \cosh \left(
	\beta h(z)
\right)
\label{TraceHamiltonianQisTwo}
\end{align}
and 
\begin{align}
\Tr{} \Spin{1}{} \exp \Hamiltonian 
= \exp \left(
	\frac{\beta^2 k}{8 \MeanK^2} ( 1-Q_{11} ) 
\right)
2 \sinh \left(
	\beta h(z)
\right),
\label{TraceSHamiltonianQisTwo}
\end{align}
where $z = z_{11}$ and $\beta h(z) = \frac{\beta(1-\eta)}{2 \MeanK} M_{1} + \frac{\beta \sqrt{k}}{2\MeanK} \sqrt{Q_{11}} z$.
Using \cref{TraceHamiltonianQisTwo,TraceSHamiltonianQisTwo}, the self-consistent equations~(\ref{SelfEqM}), (\ref{SelfEqL}), and (\ref{SelfEqQ}) become
\begin{subequations}
\begin{align}
M_{1}& = \DegreeNodeSum \MeasureReduced \tanh \left( \beta h(z) \right),  
\end{align}
\begin{align}
Q_{11}& = \DegreeNodeSum \MeasureReduced \tanh^2 \left( \beta h(z) \right), \textrm{  and}  
\end{align}
\begin{align}
L_{11}& = \DegreeNodeSum \MeasureReduced 1 = 1.
\end{align}
\label{OrderParameterQisTwo}
\end{subequations}
where $\mathcal{D} z = \frac{\dif z}{\sqrt{2 \pi}} \exp\left( -\frac{z^2}{2}  \right)$.

Finally, the free energy, \cref{FreeEnergy1}, and the energy, \cref{Energy}, become
\begin{subequations}
\begin{align}
f = &- \frac{1-\eta}{4} ( 1- M_{1}^2) 
	- \frac{\beta}{16 \MeanK} (Q_{11} - 1)^2 \nonumber \\
	&- \frac{1}{\beta} \DegreeSum \MeasureReduced \ln 2 \cosh\left(\beta h(z)\right),
\end{align}
and
\begin{align}
E/N &= - \frac{1-\eta}{4} ( 1 + M_{1}^2) 
	- \frac{\beta}{8 \MeanK} (1 - Q_{11}^2) \nonumber \\
	&= - \frac{1-\eta}{4} ( 1 + M_{1}^2) - \frac{\beta (1 + Q_{11})}{8 \MeanK} \frac{q}{\beta r \sqrt{Q_{11}}} \nonumber \\
	&\times\DegreeSum \sqrt{k} \MeasureReduced z \tanh \left(\beta h(z)\right),
\label{EnergyQisTwo}
\end{align}
\end{subequations}
respectively.
\cref{SelfEqQ} is used for deriving the last equality in the above equation.
Setting $\eta = 1$ and taking the $\beta \to \infty$ limit \footnote{Strictly speaking, the large $\beta$ limit in this study is taken while
maintaining the condition $\beta/\MeanK \ll 0$}, the ground state energy is given by
\begin{align}
E_g/N &= - \frac{1}{2 \MeanK} \DegreeSum \sqrt{k}
	\int_{-\infty}^{\infty} \frac{\dif z}
		{\sqrt{2\pi}} \exp \left(-\frac{z^2}{2} \right) \left \lvert z \right \rvert  \nonumber \\
	&= -\frac{1}{\sqrt{2\pi}} \frac{\Braket{\sqrt{k}}}{\MeanK}.
\label{EnergyQisTwoAfterLimit}
\end{align}
Note that this is in agreement with the result presented in \cite{Reichardt2007}.

\subsection{q=3}
\label{q3case}
For the $q=3$ case, $\HamiltonianComp{t}$ in \cref{HamiltonianKReduced} for each Potts spin vector can be written as
\begin{widetext}
\begin{subequations}
\begin{align}
\HamiltonianComp{1} &= 
	\frac{\beta (1-\eta) r k}{q \MeanK}  M_{1}
	+ \frac{\beta^2 r^2 k}{2 \MeanK^2 q^2}  \left(L_{11} - Q_{11} \right)
	+ \frac{\beta r}{\MeanK q} \sqrt{k Q_{11}} z_{11}, 
\end{align}
\begin{align}
\HamiltonianComp{2} &= 
	-\frac{\beta (1-\eta) r k}{2q \MeanK}  M_{1}
	+ \frac{\beta^2 r^2 k}{2 \MeanK^2 q^2} 
	\left[
		\frac{\left(L_{11} - Q_{11} \right)}{4} + \frac{3\left(L_{22} - Q_{22} \right)}{4}
	\right] 
	+ \frac{\beta r}{\MeanK q} \left[
		\frac{-\sqrt{k Q_{11}} z_{11}}{2} + \frac{\sqrt{3 k Q_{22}} z_{22} }{2}
	\right], 
\end{align}
\begin{align}
\HamiltonianComp{3} &= 
	-\frac{\beta (1-\eta) r k}{2 q \MeanK} M_{1}
	+ \frac{\beta^2 r^2 k}{2 \MeanK^2 q^2} 
	\left[
		\frac{\left(L_{11} - Q_{11} \right)}{4} + \frac{3\left(L_{22} - Q_{22} \right)}{4}
	\right] 
	+ \frac{\beta r}{\MeanK q} \left[
		 \frac{-\sqrt{k Q_{11}} z_{11}}{2} -  \frac{\sqrt{3 k Q_{22}} z_{22} }{2}
	\right].
\end{align}
\end{subequations}
\end{widetext}
Note that $\WeightSum{\Spin{\mu}{}}=\sum_{t=1}^q \Spin{t,\mu}{} \exp \HamiltonianComp{t}/\sum_{t=1}^q \exp \HamiltonianComp{t}$, where $\SpinVecComp{t}{\mu}$ is the $\mu$-th component of $\SpinVec{q}{t}$ (see \cref{appendixB}). 
As $\beta \to \infty$, the largest term in the summation dominates among $\exp \HamiltonianComp{1}$, $\exp \HamiltonianComp{2}$, and $\exp \HamiltonianComp{3}$. 
Now let us define 
\begin{enumerate}
\item $\ell_1 \equiv z_{22} - \sqrt{\frac{3 Q_{11}}{Q_{22}}} z_{11} - G$,
\item $\ell_2 \equiv z_{22} + \sqrt{\frac{3 Q_{11}}{Q_{22}}} z_{11} + G$,
\item $\ell_3 \equiv z_{22}$,
\end{enumerate}
where 
\begin{align}
G =& \frac{\sqrt{3k}(1-\eta)}{\sqrt{Q_{22}}}M_{1} \nonumber \\
	&+ \frac{\sqrt{3k}  r }{4 \MeanK q \sqrt{Q_{22}}}
	(\beta(L_{11} - Q_{11}) - \beta (L_{22} - Q_{22}) ).
\label{CrossingPoint}
\end{align}
Note that the three lines $\ell_1=0$, $\ell_2=0$, and $\ell_3=0$ meet at one point $\left( -G \sqrt{\frac{Q_{22}}{3Q_{11}}}, 0\right)$ in a 2-dimensional plane $(z_{11}, z_{22})$ and divide a whole plane into three regions $A$, $B$, and $C$ as follows:
\begin{enumerate}
\item $A : \ell_1 < 0 \quad\textrm{and}\quad \ell_2 > 0$,
\item $B : \ell_1 > 0 \quad\textrm{and}\quad \ell_3 > 0$,
\item $C : \ell_2 < 0 \quad\textrm{and}\quad \ell_3 < 0$.
\end{enumerate}
Then, one can show that $\HamiltonianComp{1}$, $\HamiltonianComp{2}$, and $\HamiltonianComp{3}$ dominate in $A$, $B$, and $C$, respectively. On these divided regions, in the $\beta \rightarrow \infty$ limit, the self-consistent equation for $M_{1}$, \cref{SelfConsistent1}, can be written in terms of the regions as 
\begin{align}
M_{1} = 1 - 3 D,
\label{magnetization}
\end{align}
where
\begin{align}
D \equiv \DegreeNodeSum \MeasureRegion{B} = \frac{1}{2} - \frac{1}{2} \DegreeNodeSum \MeasureRegion{A}.
\label{SelfD}
\end{align}
Other self-consistent equations in \cref{SelfConsistent2,SelfConsistent} can also be written in terms of the regions in the similar way. Calculation details are presented in \cref{appendixD}.
Collecting all the new self-consistent equations written in terms of the regions $A$, $B$ and $C$, the ground state energy, \cref{Energy}, can now be calculated as
\begin{align}
E_g/N =& - \frac{1-\eta}{2}(1-4D + 6D^2)  \nonumber \\
	&- \frac{1}{\MeanK} \left(
	-2 \sqrt{1 - \frac{3}{2}D} X + \sqrt{2D} Y
\right) ,
\label{EnergyGroundGeneral}
\end{align}
where
\begin{align}
X &= \DegreeSum \sqrt{k} \MeasureRegion{B} z_{11} \quad\textrm{and}
\nonumber \\
Y &= \DegreeSum \sqrt{k} \MeasureRegion{B} z_{22}.
\nonumber
\end{align}
Here, $X$ and $Y$ are the self-consistently determined quantities (see \cref{appendixD}). In the case of $\eta =1$, $D$ becomes $\frac{1}{3}$ and thus,
\begin{align}
E_g/N= -\frac{1}{\MeanK} \left(
	-\sqrt{2} X + \sqrt{\frac{2}{3}} Y
\right)
	= - \sqrt{\frac{\frac{3}{2}}{2\pi}} \frac{\Braket{\sqrt{k}}}{\MeanK}.
\label{Eforq3}
\end{align}

\subsection{q=4}
For $q=4$, calculation of the ground state energy is rather complicated and tedious, but proceeds in a similar way as the previous section; in this case, the three-dimensional plane $(z_{11}, z_{22}, z_{33})$ is divided into four regions and the self-consistent equations are written as the integral over the divided regions. 
Here, we present only the final result for the ground state energy when $\eta = 1$ as below;
\begin{widetext}
\begin{align}
E/N &= -\frac{3 \sqrt{3}}{4 \MeanK} \DegreeSum \sqrt{k} \MeasureReduced z_{\mu\mu} \frac{ \Tr{} \Spin{\mu}{} \exp \Hamiltonian}{\Tr{} \exp \Hamiltonian} \quad \quad(\textrm{for any }\mu)  \nonumber \\
&\textrm{where } \MeasureReduced z_{\mu\mu} \frac{ \Tr{} \Spin{\mu}{} \exp \Hamiltonian}{\Tr{} \exp \Hamiltonian}  
\xrightarrow{\beta \to \infty} \frac{2\sqrt{6}}{3} \int_{0}^{\infty} \dif z_{33} \int_{-\infty}^{\frac{z_3}{\sqrt{3}}} \dif z_{22} \int_{-\infty}^{\frac{\sqrt{6}z_{33}-\sqrt{2} z_{22}  }{4} } \dif z_{11}
\frac{z_{33} }{(2\pi)^{  \frac{3}{2}  }}	 \exp \left( - \frac{\sum_{\mu=1}^3 z_{\mu\mu}^2 }{2}\right) \nonumber \\
&\quad \quad \quad \quad\quad\quad\quad\quad\quad\quad\quad\quad\quad\quad\quad \approx  0.243.
\end{align}
\end{widetext}
where $\mathcal{D} z = \prod_{\mu=1}^3 \frac{\dif z_{\mu \mu}}{\sqrt{2 \pi}} \exp\left(-\frac{z_{\mu \mu}^2}{2}  \right) $.
Then, 
\begin{align}
E_g/N &\approx - \frac{3\sqrt{2}}{2} \times 0.243  \DegreeSum \sqrt{k} \nonumber \\
&\approx - \sqrt{ \frac{\frac{5}{3}}{2\pi} } \frac{\Braket{\sqrt{k}}}{\MeanK}
\label{Eforq4}
\end{align}

\subsection{Conjecture for a general $q$}
By extrapolating the results, \cref{EnergyQisTwoAfterLimit,Eforq3,Eforq4}, we conjecture a formula for the ground state energy for general $q$ with $\eta=1$ as follows;
\begin{align}
E_g/N = - \sqrt{\frac{2 - \frac{1}{q-1}}{2 \pi}} \frac{\Braket{\sqrt{k}}}{\MeanK}.
\label{Conjecture}
\end{align}
The above conjecture will be tested numerically in the next section.
Then, from \cref{QErelation}, the modularity of a random uncorrelated network with arbitrary degree distribution based on $q$ communities becomes
\begin{align}
	\left[ \ModularityMax \right]_c= 2 \sqrt{\frac{2 - \frac{1}{q-1}}{2 \pi}} \frac{\Braket{\sqrt{k}}}{\MeanK}.
	\label{modularity_solution}
\end{align}

\section{Numerical simulations}
\label{sec:Numerical}

Here, we describe the results of numerical simulations and compare them with the analytical expressions derived in the previous sections.
We used the static model introduced by Goh \emph{et al.}~\cite{Goh2001} to generate an ensemble of random networks.
The term `static' originates from the fact that the number of vertices $N$ of a network is fixed while constructing a network sample.
In this model, a normalized weight $P_i$ ($\sum_i P_i =1$) is assigned to each vertex $i$.
We consider the case whereby $P_i$ follows a power-law form, {\emph i.e.}, $P_i =i^{-\alpha}/\sum_{j}j^{-\alpha}$.
A network is constructed via the following process.
In each time step, the two vertices $i$ and $j$ are selected with
probabilities $P_i$ and $P_{j}$, respectively.
If $i=j$ or an edge connecting $i$ 
and $j$ already exists, we do nothing; otherwise, an edge is added
between vertices $i$ and $j$.
We repeat this step $NK$ times.
The probability that a given pair of vertices $i$ and
$j (i\neq j)$ are not connected by an edge following this process is given by $(1-2P_{i}P_{j})^{NK}\simeq e^{-2NKP_{i}P_{j}}$.
Thus, the connection probability for nodes $i$ and $j$ is $1-e^{2NKP_{i}P_{j}}$.
Here, we used the condition $P_{i}\ll 1$.
The factor 2 in the exponent comes from the equivalence of $(ij)$ and $(ji)$.
The connection probability $f_{ij}$ can thus be approximated as $f_{ij} \approx 2NKP_{i}P_{j}
	\approx \langle k_i \rangle \langle k_j \rangle/(\MeanK N) $ in the thermodynamic limit, 
where we used the fact $\langle k_i \rangle =
2KNP_i$ in this limit~\cite{Lee2006}.
The resulting network is scale-free and has a degree exponent $\gamma$ given by 
\begin{align}
\gamma = 1+\frac{1}{\alpha}.
\end{align}
Note that a network generated by the static model becomes uncorrelated when $\gamma\geq 3$~\cite{Lee2006}. Therefore, we performed the simulation on a network with $\gamma\geq 3$.
For this scale-free network, the \cref{modularity_solution} becomes
\begin{align}
	\left[\ModularityMax \right]_c = 2 \sqrt{\frac{2 - \frac{1}{r}}{2 \pi}} \frac{\sqrt{(\gamma-1)(\gamma-2)}}{(\gamma-\frac{3}{2})} \MeanK^{-1/2}.
	\label{SF_modularity}
\end{align}
The size of the networks $N$ used in this study was $10000$, and the exponents of the degree distributions were $3.0$, $3.5$, $4.0$, and $4.5$.
As $\gamma \to \infty$ limit, we also performed the same numerical simulations for the Erd\H{o}s-R\'{e}nyi (ER) network \cite{Erdos1959} of the same size.

Since finding the ground state of the Potts model Hamiltonian is an NP-hard problem, it is practically impossible to do so for very large networks.
Instead, we used the simulated annealing method~\cite{Kirkpatrick1983} to obtain an approximate solution.
Initially, one of the $q$ possible spins was randomly assigned to each node in the network. The initial temperature was set to be sufficiently high.
In the Monte Carlo simulation, we chose one spin at random, and determined whether the spin state was changed according to the Metropolis algorithm.
This procedure was repeated until the system reached a stationary state at a fixed temperature.
The temperature was then reduced according to a predefined schedule, and the simulation was repeated until it reached a stationary state for this new temperature.
The final state, {\emph i.e.}, the stationary state at zero temperature, was assumed to be the ground state of the system.

\begin{figure}
\centering
\includegraphics[width=0.95\columnwidth]{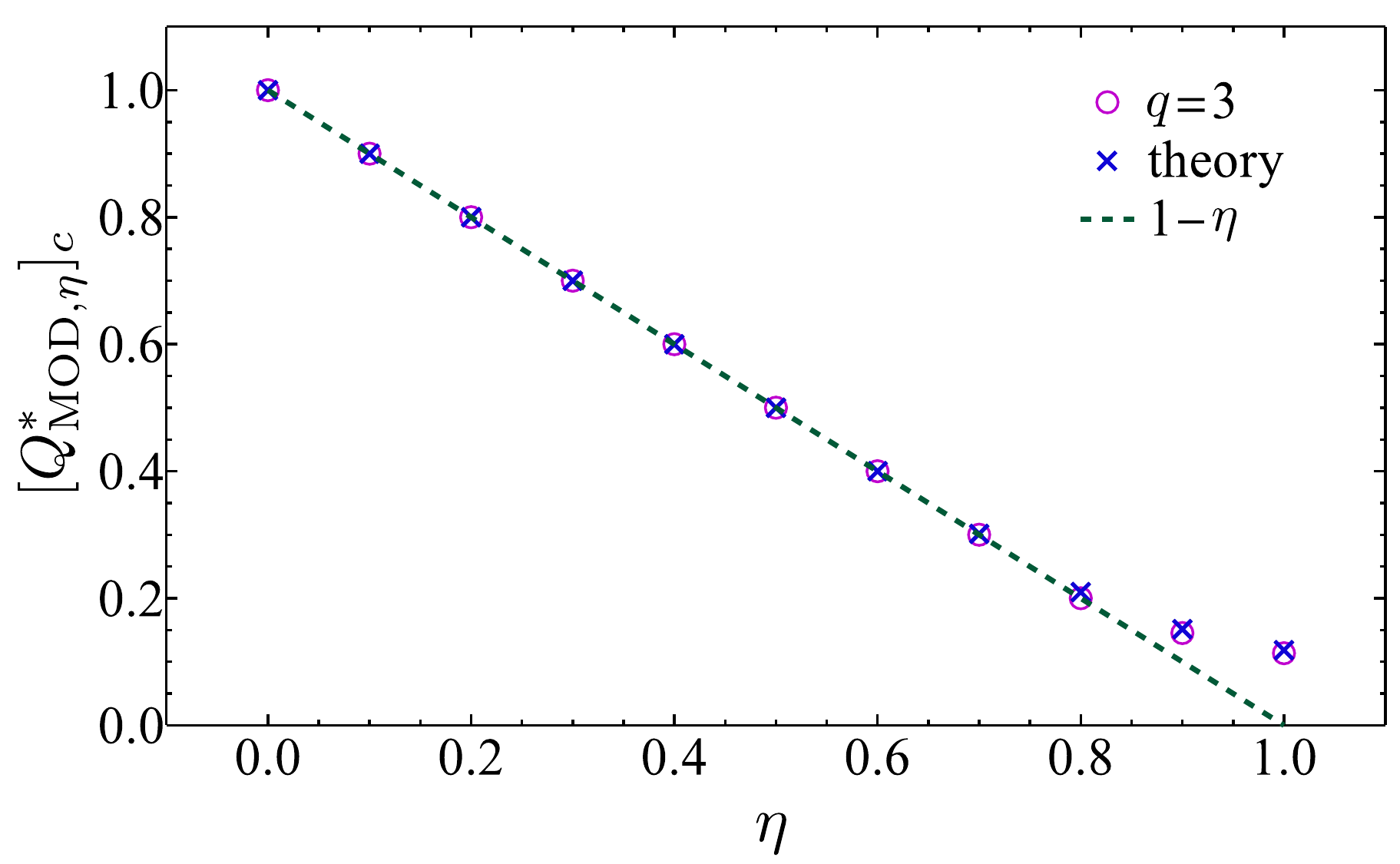}
\caption{(color online) Plot of $\ModularityMaxEta$ against $\eta$ for $q=3$, $N=10000$, $\MeanK=64$, and $\gamma=3.5$.
The blue open circles represent the data calculated using the simulated annealing method. 
The cross symbols indicate the solutions of \cref{EnergyGroundGeneral} obtained by solving the self-consistent equations (\ref{CrossingPoint}), (\ref{SelfD}), (\ref{SelfX}), and (\ref{SelfY}) numerically.
For small $\eta$, $D$ is expected to be small, and thus $\ModularityMaxEta$ is very close to $1-\eta$ (the red dashed curve), can be see in \cref{EnergyGroundGeneral}.
}
\label{fig:EtaDependence}
\end{figure}

\cref{fig:EtaDependence} shows a plot of $\ModularityMaxEta$ versus $\eta$ for $q=3$, $\MeanK=64$, and $\gamma=3.5$.
The analytical results were in very good agreement with the simulated data. 
As $\eta$ approached $0$, the interaction between Potts spins became more ferromagnetic and $M_1 \to 1$.
We then found that $D \to 0$ from \cref{magnetization}, which made $\ModularityMaxEta \approx 1-\eta$, from \cref{EnergyGroundGeneral}.

\begin{figure}
\includegraphics[width=0.95\linewidth]{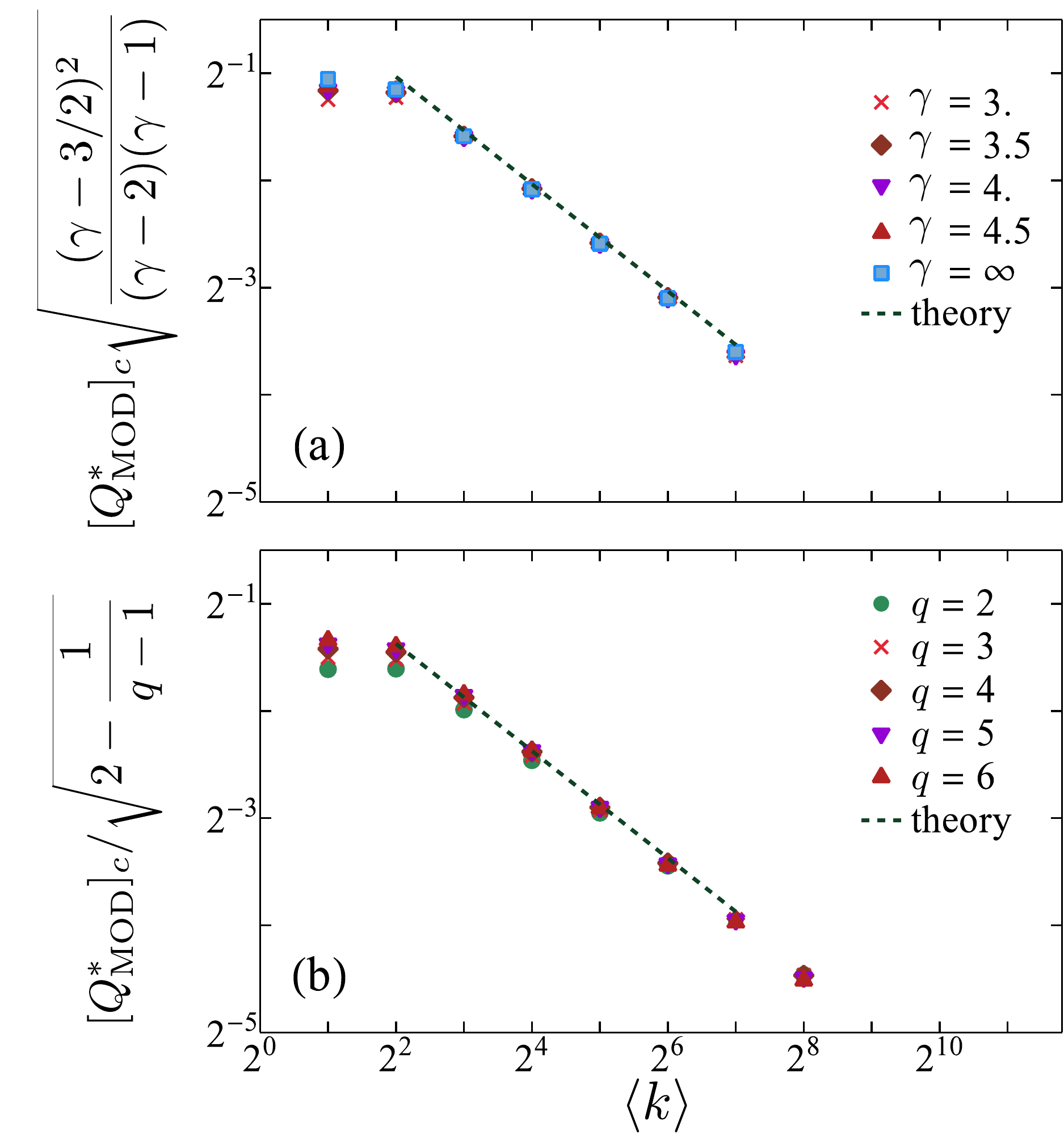}
\caption{(a) Plot of the rescaled $\ModularityMaxAvr$ as a function of $\MeanK$ for 
$\gamma= 2.5$, 3, 3.5, 4, 4.5, and
ER ($\gamma \to \infty$), with $\eta=1.0$ and $q=3$.
The red dashed curve shows the result of \cref{SF_modularity}.
The gradient of the curve in the double logarithmic scale was $-0.5$.
(b) Plot of the rescaled $\ModularityMaxAvr$ as a function of $\MeanK$ for 
$q=2$, 3, 4, 5, and 6 with $\eta=1.0$ and $\gamma=3.5$.
The collapsing behavior of the data indicates the validity of the conjecture, \cref{modularity_solution} or \cref{SF_modularity}. The red dashed curve shows the theoretically conjectured curve from \cref{SF_modularity}.
} \label{fig:QGammaDependence}
\end{figure}

\cref{fig:QGammaDependence} (a) shows $\ModularityMaxAvr$ as a function of $\MeanK$ for various $\gamma$ with $\eta=1.0$ and $q=3$.
Note that $\ModularityMaxAvr$ was rescaled by $\frac{\sqrt{(\gamma-1)(\gamma-2)}}{(\gamma-3/2)}$ in order to observe the collapsing behavior.
As can be expected from the analytical results, all the simulated data collapsed onto the curve given by \cref{SF_modularity}.
\cref{fig:QGammaDependence} (b) shows $\ModularityMaxAvr$ as a function of $\MeanK$ for various $q$ with $\eta=1.0$ and $\gamma=3.5$. In this case, $\ModularityMaxAvr$ was rescaled by $\sqrt{ 2-\frac{1}{q-1} }$. This collapsing behavior confirms our conjecture (\ref{Conjecture}) for the ground state energy of the Potts model for $q>4$.
The correspondence between our theoretical and simulated data indicates that the replica symmetric (RS) solution is valid for calculating the energy of the Potts model. We also note that the analytical results can be improved by taking into account the replica symmetry breaking (RSB) solutions. For example, as stated in Ref.~\cite{Reichardt2007}, for $q=2$, the modularity obtained from the RSB solution is more accurate. The difference in modularity between RS and RSB was approximately 6\%. However, this small difference is not significant in the logarithmic scale, as can be seen from \cref{fig:QGammaDependence}.

\section{Conclusion}

We have described a community detection method based on maximizing the modularity function, which is equivalent to finding a ground state energy of the $q$-state Potts model Hamiltonian, \cref{PottsH}, when $\eta=1$. Because a random uncorrelated network has a finite modularity due to quenched disorder, the modularity of a given network is meaningful only when it is compared with that of a random network. Therefore, we analytically calculated the modularity of a random uncorrelated network as a reference by finding the ground state energy of the $q$-state Potts model. We used the replica method find a replica symmetric solution. We also studied the densely connected regime where $\beta \ll \langle k \rangle$, even if we take the limit $\beta \rightarrow \infty$, which is described formally at the later stages of the calculation.

We showed that, for an arbitrary $q$, the modularity is proportional to $\MeanK^{-0.5}$ when $\eta=1$ in the large average degree limit. 
We also performed simulations using the simulated annealing method to find the ground state of the $q$-state Potts model and showed that our analytical results were in good agreement with the simulated data.
Our results provide a theoretical minimum value over which the modularity of a network becomes meaningful. In addition, our calculation method may be applicable to evaluating the energy of a similar type of $q$-state Potts model.

\begin{acknowledgments}
This research was supported by the NRF grant Nos.~2011-35B-C00014 (JSL) and 2010-0015066 (BK).
\end{acknowledgments}

\appendix


\section{Linearization of quadratic spin product}
\label{appendix:Hubbard}

We begin by introducing the modified Hubbard-Stratonovich transform as follows;
\begin{align}
\exp(2 \lambda^2 a b) &= 
\iint \frac{\dif z}{\sqrt{2\pi}} \frac{\dif w}{\sqrt{2\pi}} \nonumber \\
&\times  \exp \left( 
	-\frac{1}{2}z^2 -\frac{1}{2} w^2+ \lambda (a+b) z + i \lambda (a-b) w
\right).
\end{align}
Using the above transformation, the last term of the exponent in \cref{replicaSymHamiltonian} becomes,
\begin{widetext}
\begin{align}
&\exp \left[
	\frac{k C_2}{\MeanK}  \sum_{\mu\nu} Q_{\mu \nu}
	\left( 
		\sum_{\alpha } \Spin{\mu}{\alpha}  \right) \left( \sum_{\beta } \Spin{\nu}{\beta} 
	\right)
\right]  \nonumber \\
&= \Measure  \exp 
\left[
	\sum_{\mu\nu}\sum_{\alpha} \sqrt{\frac{k C_2 Q_{\mu \nu}}{2\MeanK}} 
	\Set{
		\left(  \Spin{\mu}{\alpha} + \Spin{\nu}{\alpha} \right) z_{\mu \nu}
		+ i \left(  \Spin{\mu}{\alpha} - \Spin{\nu}{\alpha} \right) w_{\mu \nu}
	}
\right],
\end{align}
\end{widetext}
where $\mathcal{D} z \mathcal{D}  w  = \prod_{\mu\nu} \frac{\dif z_{\mu \nu}}{\sqrt{2 \pi}}\frac{\dif w_{\mu \nu}}{\sqrt{2 \pi}} \exp\left(-\frac{z_{\mu \nu}^2}{2} -\frac{w_{\mu \nu}^2}{2} \right) $.
Note that the term quadratically coupled by two replica indices is now linearized in the final expression.
Then, the trace of $\exp \HamiltonianFull$ of \cref{replicaSymHamiltonian} can be evaluated as,
\begin{align}
\Tr{\alpha}  \exp  \HamiltonianFull  = \Measure \left(
	\Tr{} \exp \Hamiltonian
\right)^n,
\end{align}
where $\Hamiltonian$ is defined in \cref{HamiltonianK}.


\section{Vector representation of $q$-states Potts model}
\label{appendixB} 
Consider a $r$-dimensional simplex with $q $ vertices whose center of mass is located at the origin.  
If we define $\theta_q$ be the angle between any two vectors pointing from the origin to the vertices of the simplex, it satisfies $\cos \theta_{q} = -\frac{1}{q-1}$. Because Potts spin vectors can be identically mapped to the vectors of the simplex~\cite{Wikipedia2013}, $r$-dimensional Potts vector can be expressed by $\theta_q$.
For $q=2$, $\SpinVec{2}{1}=(1)$ and $\SpinVec{2}{2}=(\cos \theta_2)$.
For $q=3$, $\SpinVec{3}{1}=(1, 0)$, $\SpinVec{3}{2}=(\cos\theta_3, \sin\theta_3)$ and $\SpinVec{3}{3}=(\cos\theta_3, \sin\theta_3 \cos \theta_2)$. 
Apart from $\SpinVec{3}{1}$, the other two vectors can be written as, $\SpinVec{3}{2}= \cos \theta_3 || \sin \theta_3 \SpinVec{2}{1}$ and $\SpinVec{3}{3} = \cos \theta_3 || \sin \theta_3 \SpinVec{2}{2}$, where the concatenation operator $||$ is defined as $a || (b_1, b_2, \cdots, b_\ell) \equiv (a, b_1, b_2, \cdots, b_\ell)$. 
With this operator, the $q$-states Potts spin vectors can be written as $\SpinVec{q}{1}=(1, 0, \cdots, 0)$ and $\SpinVec{q}{\ell} = \cos \theta_{q} || \sin \theta_{q} \SpinVec{q-1}{\ell-1}$ for $\ell = 2, \cdots, q$.
By construction, one can prove the several identities stated below.
Let $\SpinVecComp{t}{\mu}$ be the $\mu$-th element of $\SpinVec{q}{t}$. 
Then, one can verify 
\begin{align}
\SpinVecComp{t}{\mu} = 0,
\label{SpinVecIdentity1}
\end{align}
for $1 \le t \le \mu - 1$   
and 
\begin{align}
\SpinVecComp{t}{\nu} = \SpinVecComp{\nu+1}{\nu},
\label{SpinVecIdentity2}
\end{align}
for $\nu < t$.
It can also be shown that 
\begin{subequations}
\begin{align}
\sum_{\mu=1}^{q-1} \SpinVecComp{t}{\mu} \SpinVecComp{u}{\mu} &= 1-(1-\delta_{t u})(1-\cos\theta_q) , \label{SumRuleOne}
\end{align}
\begin{align}
\sum_{t=1}^{q} (\SpinVecComp{t}{\mu})^2 &= \frac{q}{q-1},\label{SumRuleTwo}
\end{align}
\begin{align}
\sum_{t=1}^{q} \SpinVecComp{t}{\mu} &= 0. \label{SumRuleThree}
\end{align} 
\end{subequations}

\section{Properties of $L_{\mu \nu}$ and $Q_{\mu \nu}$}
\label{appendixC}
In \cref{SelfEqL}, the expression $\WeightSum{\Spin{\mu}{}\Spin{\nu}{}}$ for $\mu > \nu$ can be simplified as,
\begin{align}
\WeightSum{\Spin{\mu}{}\Spin{\nu}{}} &= \frac{\Tr{} \Spin{\mu}{}\Spin{\nu}{} \exp \Hamiltonian}{\Tr{} \exp \Hamiltonian} \nonumber \\
&= \sum_{t=1}^{q} \frac{\SpinVecComp{t}{\mu} \SpinVecComp{t}{\nu} \exp \HamiltonianComp{t}}{\Tr{} \exp \Hamiltonian} \nonumber \\
&= \sum_{t=\mu}^{q} \frac{\SpinVecComp{t}{\mu} \SpinVecComp{t}{\nu} \exp \HamiltonianComp{t}}{\Tr{} \exp \Hamiltonian} \nonumber \\
&= \SpinVecComp{\mu}{\nu} \sum_{t=\mu}^{q} \frac{\SpinVecComp{t}{\mu}  \exp \HamiltonianComp{t}}{\Tr{} \exp \Hamiltonian} \nonumber \\
&= \SpinVecComp{\mu}{\nu} \WeightSum{\Spin{\mu}{}},
\end{align}
where $\HamiltonianComp{t}$ denotes a $\Hamiltonian$ calculated at $\SpinVec{q}{t}$. 
Note that \cref{SpinVecIdentity1,SpinVecIdentity2} are used for the third and fourth equalities, respectively, in the above equation. 
Using the facts that $\Average{\Spin{\mu}{}} = M_{\mu} = 0$ for $\mu > 1$ and $L_{\mu\nu} = L_{\nu \mu}$,
we have
\begin{align}
L_{\mu\nu} = 0
\label{ConditionL}
\end{align} 
for $\mu \ne \nu$.

Next, we will show that $L_{\mu \mu} = L_{\nu \nu}$ for $\mu > 1$ and $\nu > 1$.
First, it is useful to consider the sum of $L_{\mu \mu}$.
\begin{align}
\sum_{\mu=1}^{q} L_{\mu \mu}
&= \sum_{\mu=1}^{q} \Average{ \Spin{\mu}{2} } \nonumber \\
&= \Average{\sum_{\mu=1}^{q} \Spin{\mu}{2}} = 1,
\label{SumOrderL}
\end{align}
where \cref{SumRuleOne} is used for the last equality. Here, $\mathcal{D} z \mathcal{D}  w  = \prod_{\mu\nu} \frac{\dif z_{\mu \nu}}{\sqrt{2 \pi}}\frac{\dif w_{\mu \nu}}{\sqrt{2 \pi}} \exp\left(-\frac{z_{\mu \nu}^2}{2} -\frac{w_{\mu \nu}^2}{2} \right) $.
To proceed further, we define a quantity,  
\begin{align}
V_t \equiv \DegreeNodeSum \Measure \frac{\exp \HamiltonianComp{t}}{\Tr{} \exp \Hamiltonian}.
\end{align}
Then, $M_\mu$ can be written as the sum of $V_t$ and $ \SpinVecComp{t}{\mu} $, i.e., 
\begin{align}
M_\mu = \sum_{t=1}^{q}  V_t \SpinVecComp{t}{\mu},
\label{LinearEq}
\end{align}
for $\mu = 1, \cdots, q-1$.
The set of linear equations, \cref{LinearEq}, can be solved and the solution is
\begin{align}
V_m = V_2
\label{IdentityV}
\end{align}
for $m = 3, \cdots, q$ and $V_1 = M_{1} + V_2$.
Thus, one finds that for $\mu > 1$,
\begin{align}
L_{\mu \mu} 
&= \Average{\Spin{\mu}{2}} \nonumber \\ 
&= \sum_{t=1}^{q} V_t \SpinVecComp{t}{\mu}^2 = \sum_{t=\mu}^{q} V_t \SpinVecComp{t}{\mu}^2 \nonumber \\
&= V_2 \sum_{t=\mu}^{q} \SpinVecComp{t}{\mu}^2 = V_2 \frac{q}{q-1},
\label{OrderParamLIdentity}
\end{align}
where \cref{SpinVecIdentity1,IdentityV,SumRuleTwo} are used for the third, last and fourth equalities, respectively.
Similarly, we obtain
\begin{align}
L_{11} = V_1 + \frac{V_2}{q-1}.
\label{OrderParamLIdentitySpecial}
\end{align}
Plugging \cref{OrderParamLIdentity,OrderParamLIdentitySpecial} into \cref{SumOrderL}, 
	$L_{\mu\mu}$ is given by,
\begin{subequations}
\begin{align}
L_{11} = \frac{1 + (q-2)M_1}{q-1}
\label{OrderLFinFirst}
\end{align}
and
\begin{align}
L_{22} = \frac{1 - M_1}{q-1}=L_{\nu\nu} ~~~\textrm{for }\nu>1.
\label{OrderLFin}
\end{align}
\end{subequations} 

Next, let us examine the properties of $Q_{\mu\nu}$. 
We will show that there exist non-trivial solutions for the self-consistent \cref{SelfEqQ}
	satisfying the following conditions 
\begin{subequations}
\begin{align}
Q_{\mu \nu} = 0
\label{ConditionQ} 
\end{align}
for $\mu \ne \nu$ and 
\begin{align} Q_{\mu \mu} = Q_{\nu \nu}
\label{ConditionInvariance}
\end{align}
for $\mu>1$ and $\nu >1$.
\label{ConditionAll}
\end{subequations}
With these conditions, $\HamiltonianComp{t}$ in \cref{HamiltonianK} can be written as
\begin{align}
\HamiltonianComp{t} &\equiv 
	\frac{k C_1}{\MeanK}  \Spin{t,1}{} M_{1} \nonumber \\
	&+ \frac{k C_2}{\MeanK}\Spin{t,1}{2} \left(L_{11} - Q_{11} \right) 
	+\frac{k C_2}{\MeanK} \left(L_{22} - Q_{22} \right) \sum_{\mu=2}^r \Spin{t,\mu}{2}  \nonumber \\
	&+ \sqrt{\frac{2 k C_2 Q_{11}}{\MeanK}} \Spin{t,1}{} z_{11}  
	+ \sqrt{\frac{2 k C_2 Q_{22}}{\MeanK}} \sum_{\mu = 2}^{r} \Spin{t,\mu}{} z_{\mu\mu}.
\label{HamiltonianKReduced}
\end{align}
If we define $\widetilde{Q}_{11} = \sqrt{\frac{2 k C_2 Q_{11}}{\MeanK}} $ and $\widetilde{Q}_{22} = \sqrt{\frac{2 k C_2 Q_{22}}{\MeanK}} $, $\HamiltonianComp{t}$ for $t=1$ and $t=w>1$ can be written as
\begin{align}
\HamiltonianComp{1} &= 
	\mathcal{B}_1 + \widetilde{Q}_{11} z_{11}, \textrm{ and}
	 \nonumber \\
\HamiltonianComp{w} &=
	\mathcal{B}_2 - \widetilde{Q}_{11} \frac{z_{11}}{r}  
	+ \widetilde{Q}_{22}   \SpinVec{q}{w}\cdot \vec{z^\prime},
\label{HamiltonianKSpecific}
\end{align}
respectively, where 
\begin{align}
\mathcal{B}_1&= \frac{k C_1}{\MeanK}  M_{1} + \frac{k C_2}{\MeanK} \left(L_{11} - Q_{11} \right), \nonumber \\
\mathcal{B}_2&= -\frac{k C_1}{\MeanK}  \frac{M_{1}}{r} \nonumber \\ &+ \frac{k C_2}{r^2\MeanK} \left(L_{11} - Q_{11} \right) 
	+\frac{k C_2 (r^2-1)}{\MeanK r^2} \left(L_{22} - Q_{22} \right), \nonumber \\ 
\SpinVec{q}{t} &= (\Spin{t,1}{},\Spin{t,2}{},\cdots,\Spin{t,r}{}), \nonumber \\
\vec{z^\prime} &= (0,z_{22},z_{33},\cdots,z_{rr}).
\end{align}
Note that $\mathcal{B}_1$ and $\mathcal{B}_2$ have nothing to do with the auxiliary integration variables $z_{\mu\mu}$.

From \cref{SelfEqQ}, $Q_{\mu\mu}$ can be written as
\begin{align}
Q_{\mu\mu} &=\DegreeNodeSum \MeasureReduced \frac{\left( \sum_{t=1}^q \Spin{t,\mu}{} \exp \HamiltonianComp{t} \right)^2}{ \left( \sum_{t=1}^q \exp \HamiltonianComp{t}  \right)^2 } \nonumber \\
&=\sum_{u,v} \Spin{u,\mu}{}\Spin{v,\mu}{} \DegreeNodeSum \MeasureReduced \frac{\exp \HamiltonianComp{u}  \exp \HamiltonianComp{v}}{ \left( \sum_{t=1}^q \exp \HamiltonianComp{t} \right)^2 }.
\label{h_t_h_t}
\end{align}
Note that in the above equation the integral with respect to $\int\mathcal{D} w$ disappears because $\SpinAsymm =0$ for $\nu=\mu$ and $Q_{\mu\nu}=0$ for $\nu\neq\mu$,
thus, all the integration variables $w_{\mu\nu}$ in $\HamiltonianComp{}$ in \cref{HamiltonianK} vanish.
In addition, now $\mathcal{D} z = \prod_{\mu} \frac{\dif z_{\mu \mu}}{\sqrt{2 \pi}} \exp\left(-\frac{z_{\mu \mu}^2}{2}  \right) $ because the off-diagonal terms of $ z_{\mu \nu}$ also vanish by \cref{ConditionQ}.
Then, the integral in \cref{h_t_h_t} can be categorized into the following four cases.
\begin{align}
& \MeasureReduced \frac{ \exp \HamiltonianComp{u} \exp \HamiltonianComp{v}}{ \left( \sum_{t=1}^q \exp \HamiltonianComp{t} \right)^2 } \nonumber \\
= & \begin{dcases}
	\MeasureReduced \frac{ \exp \HamiltonianComp{1} \exp \HamiltonianComp{1}}{ \left( \sum_{t=1}^q \exp \HamiltonianComp{t} \right)^2 },
		& \For \quad u=v=1  \\ 
	\MeasureReduced \frac{ \exp \HamiltonianComp{1} \exp \HamiltonianComp{2}}{ \left(\sum_{t=1}^q \exp \HamiltonianComp{t} \right)^2 }, 
		& \For \quad u=1 \textrm{ and } v > 1 \\
		\MeasureReduced \frac{ \exp \HamiltonianComp{2} \exp \HamiltonianComp{2}}{ \left( \sum_{t=1}^q \exp \HamiltonianComp{t}\right)^2 }, 
		& \For \quad u= v > 1 \\
		\MeasureReduced \frac{ \exp \HamiltonianComp{2} \exp \HamiltonianComp{3}}{ \left(\sum_{t=1}^q \exp \HamiltonianComp{t} \right)^2 }.
		& \For \quad u,v >1 \textrm{ and } u\neq v
\label{IdentityOrderQ}
\end{dcases}
\end{align}
The derivation for the above equation is straightforward. For example, for $u=1$ and $v > 1$ (the second case), using \cref{HamiltonianKSpecific}, the integral becomes
\begin{align}
\MeasureReduced \frac{ e^{\HamiltonianComp{1}+\mathcal{B}_2-\widetilde{Q}_{11}\frac{z_{11}}{r}} e^{\widetilde{Q}_{22} \SpinVec{q}{v}\cdot \vec{z^\prime} } }{ \left(e^{\HamiltonianComp{1}} + e^{\mathcal{B}_2-\widetilde{Q}_{11}\frac{z_{11}}{r}} \sum_{t=2}^q e^{ \widetilde{Q}_{22} \SpinVec{q}{t}\cdot \vec{z^\prime} } \right)^2 }.
\label{integralInvariance}
\end{align}
Because $\SpinVec{q}{t}$ for $t > 1$ possesses rotational symmetry in the subspace spanned by $z_{22}$, $z_{33}$, $\cdots$, and $z_{rr}$, \cref{integralInvariance} is invariant under the exchange of different $v(>1)$.
Therefore, the integral is the same as the integral for $u=1$ and $v=2$.
The other cases can be derived in the similar way. 

From \cref{h_t_h_t,IdentityOrderQ}, $Q_{11}$ becomes
\begin{widetext}
\begin{align}
Q_{11}
	&=\sum_{u} \Spin{u,1}{2} \DegreeNodeSum \MeasureReduced \frac{\exp 2h_u^{(k)} }{ \left( \sum_{t=1}^q \exp \HamiltonianComp{t} \right)^2 } +\sum_{u\neq v} \Spin{u,1}{} \Spin{v,1}{} \DegreeNodeSum \MeasureReduced \frac{\exp \HamiltonianComp{u} \exp \HamiltonianComp{v}}{ \left( \sum_{t=1}^q \exp \HamiltonianComp{t} \right)^2 } \nonumber \\
	&=\Spin{1,1}{2}\DegreeNodeSum \MeasureReduced \frac{\exp 2h_1^{(k)} }{ \left( \sum_{t=1}^q \exp \HamiltonianComp{t} \right)^2 } +\sum_{u=2}^{q}\Spin{u,1}{2}\DegreeNodeSum \MeasureReduced \frac{\exp 2h_2^{(k)} }{ \left( \sum_{t=1}^q \exp \HamiltonianComp{t} \right)^2 } \nonumber \\
	&+2\sum_{v=2}^q \Spin{1,1}{} \Spin{v,1}{} \DegreeNodeSum \MeasureReduced \frac{\exp \HamiltonianComp{1} \exp \HamiltonianComp{2}}{ \left( \sum_{t=1}^q \exp \HamiltonianComp{t} \right)^2 } 
	+\left[ \left(\sum_{u=2} \Spin{u,1}{}\right)^2 -\sum_{u=2}^q \Spin{u,1}{2}\right]  \DegreeNodeSum \MeasureReduced \frac{\exp \HamiltonianComp{2} \exp \HamiltonianComp{3}}{ \left( \sum_{t=1}^q \exp \HamiltonianComp{t} \right)^2 } \nonumber \\
	&=\DegreeNodeSum \MeasureReduced \left[ \frac{\exp 2h_1^{(k)} +\frac{1}{r}\exp 2h_2^{(k)} -2\exp h_1^{(k)}\exp h_2^{(k)} -(1-\frac{1}{r})\exp h_2^{(k)}\exp h_3^{(k)}}{ \left( \sum_{t=1}^q \exp \HamiltonianComp{t} \right)^2 } \right].
\label{Q_11_final}
\end{align}

For the third equality, we used \cref{SumRuleTwo,SumRuleThree}.
Using the similar way, we can find $Q_{\mu\mu}$ as
\begin{align}
Q_{\mu \mu} &= \frac{r+1}{r} \DegreeNodeSum \MeasureReduced \left(
	\frac{ \exp \HamiltonianComp{2} \exp \HamiltonianComp{2}}{ \left( \sum_{t=1}^q \exp \HamiltonianComp{t} \right)^2 }
	- \frac{ \exp \HamiltonianComp{2} \exp \HamiltonianComp{3}}{\left( \sum_{t=1}^q \exp \HamiltonianComp{t} \right)^2 }
\right)=Q_{\nu\nu},
\label{consistentSolution1}
\end{align}
\end{widetext}
for all $\mu,\nu>1.$
Finally, we can also check that  
\begin{align}
Q_{\mu \nu} &= \DegreeNodeSum \MeasureReduced \WeightSum{ \Spin{\mu}{} }  \WeightSum{ \Spin{\nu}{} } =0.
\label{consistentSolution2}
\end{align}
for all $\mu \neq \nu$ pairs. 
\cref{consistentSolution1,consistentSolution2} consistently satisfy the initially imposed conditions, \cref{ConditionQ,ConditionAll}.
Even though it is not clear whether there exist another solutions for $Q_{\mu\nu}$ from the self-consistent equations which do not satisfy \cref{ConditionAll},
	these imposed conditions must be satisfied in the $\beta \rightarrow \infty$ limit.
From \cref{SelfEqQ}, we can see that $\beta(L_{\mu\nu}-Q_{\mu\nu})$ remains finite as $\beta \rightarrow \infty$,
	which indicates $Q_{\mu\nu} \rightarrow L_{\mu\nu}$ in the zero temperature limit.
Note that $L_{\mu\nu}$ satisfies $L_{\mu\nu}=0$ for $\mu\neq\nu$ and $L_{\mu\mu}=L_{\nu\nu}$ for $\mu,\nu>1$ (see \cref{ConditionL,OrderLFin}).
Therefore, it is reasonable to impose the same conditions for $Q_{\mu\nu}$ at least in the large $\beta$ limit.

Finally, we briefly discuss on the properties of $Q_{\mu\nu}$ when $\eta=1$. In this case, using the similar method presented above, we can show that there exist non-trivial solutions for the self-consistent equations~(\ref{SelfEqQ}) satisfying \cref{ConditionQ} for $\mu\neq\nu$ and 
\begin{align}
Q_{\mu\mu}&=\frac{r+1}{r} \DegreeNodeSum \nonumber \\
 &\times\MeasureReduced \left(
	\frac{ \exp \HamiltonianComp{1} \exp \HamiltonianComp{1}}{ \left( \sum_{t=1}^q \exp \HamiltonianComp{t} \right)^2 }
	- \frac{ \exp \HamiltonianComp{1} \exp \HamiltonianComp{2}}{\left( \sum_{t=1}^q \exp \HamiltonianComp{t} \right)^2 }
\right) \nonumber \\
 &=Q_{\nu\nu},
\label{eta1caseConditionQ}
\end{align}
for all $\mu$ and $\nu$.

\section{Calculation details for the ground state energy with $q=3$}
\label{appendixD}

Since $\exp\HamiltonianComp{1}$, $\exp\HamiltonianComp{2}$ and $\exp\HamiltonianComp{3}$ dominate in the regions $A$, $B$ and $C$ (defined in \cref{q3case}), respectively, in the $\beta \rightarrow \infty$ limit, the self-consistent equation for $M_{1}$, \cref{SelfConsistent1}, can be written as 
\begin{widetext}
\begin{align}
M_{1} = \DegreeNodeSum \left(
	\MeasureRegion{A} + \MeasureRegion{B} \left( -\frac{1}{2} \right) + \MeasureRegion{C} \left( -\frac{1}{2} \right)
\right)
\end{align}
where $\MeasureRegion{R}$ denotes the integral over the domain $R$ and $\mathcal{D} z = \prod_{\mu=1}^2 \frac{\dif z_{\mu \mu}}{\sqrt{2 \pi}} \exp\left(-\frac{z_{\mu \mu}^2}{2}  \right) $.
Note that $\MeasureRegion{B} = \MeasureRegion{C}$ by symmetry and $\MeasureRegion{A \cup B \cup C} = 1$. 
\end{widetext}
From these identities, one can show that 
\begin{align}
D \equiv \DegreeNodeSum \MeasureRegion{B} = \frac{1}{2} - \frac{1}{2} \DegreeNodeSum \MeasureRegion{A}.
\label{SelfD}
\end{align}
The magnetization $M_1$, thus, can be written in terms of $D$ as, 
\begin{align}
M_{1} = 1 - 3 D.
\label{magnetization}
\end{align}
Similarly, from \cref{SelfConsistent2,SelfConsistent}, one can find
\begin{align}
L_{11} &= Q_{11} = 1 - \frac{3}{2} D
\end{align}
and
\begin{align}
L_{22} &= Q_{22} = \frac{3}{2} D.
\end{align}
To obtain the ground state energy, we should calculate $\beta\left( L_{\mu\mu} -Q_{\mu\mu} \right)$ in the $\beta \to \infty$ limit.
From the first equality in \cref{SelfConsistent3}, one can obtain
\begin{align}
\beta \left( L_{11} - Q_{11} \right) &= \frac{-3q}{r \sqrt{Q_{11}}}  X,
\end{align}
where 
\begin{align}
X \equiv \DegreeSum \sqrt{k} \MeasureRegion{B} z_{11}.
\label{SelfX}
\end{align}
For the derivation of the above equation, we used the facts, $\MeasureRegion{A \cup B \cup C} z_{11} =0$ and $\MeasureRegion{B} z_{11} = \MeasureRegion{C} z_{11}$.
Similarly, one can show that
\begin{align}
\beta \left( L_{22} - Q_{22} \right) 
	&= \frac{\sqrt{3}q}{r \sqrt{Q_{22}}} Y,
\end{align}
where 
\begin{align}
Y \equiv \DegreeSum \sqrt{k} \MeasureRegion{B} z_{22}
\label{SelfY}
\end{align}
using the identity $\MeasureRegion{B} z_{22} = - \MeasureRegion{C} z_{22}$.

Note that the following three facts: i) $D$, $X$ and $Y$ are determined by the region $B$, ii) the region $B$ depends on $G$, and iii) $G$ is evaluated from $D$, $X$ and $Y$.
Therefore, the equation \eqref{CrossingPoint} and the set of equations \eqref{SelfD}, \eqref{SelfX} and \eqref{SelfY} form self-consistent equations.

\bibliography{QStatePottsModel}

\end{document}